\documentclass[12pt]{article}
\usepackage{amsfonts,amssymb,amsmath,mathrsfs}
\usepackage{hyperref}
\usepackage[small]{caption2}
\newcommand{\beq}{\begin{equation}}
\newcommand{\eeq}{\end{equation}}
\newcommand{\vp}{\vphantom}

\newcommand{\wt}{\widetilde}

\newcommand{\pt}{\partial}
\newcommand{\al}{\alpha}
\newcommand{\bt}{\beta}

\newcommand{\de}{\delta}
\newcommand{\varep}{\varepsilon}
\newcommand{\s}{\sigma}
\newcommand{\mc}{\mathcal}

\newcommand{\un}{\underline}

\begin{document}
\begin{center}
{\Large\textbf{Ambitwistors, oscillators and massless fields on $AdS_5$}}\\[0.3cm]
{\large D.V.~Uvarov\footnote{E-mail: d\_uvarov@\,hotmail.com}}\\[0.2cm]
\textit{NSC Kharkov Institute of Physics and Technology,}\\ \textit{61108 Kharkov, Ukraine}\\[0.5cm]
\end{center}
\begin{abstract}
Positive energy unitary irreducible representations of $SU(2,2)$
can be constructed with the aid of bosonic oscillators in
(anti)fundamental representation of $SU(2)_L\times SU(2)_R$ that
are closely related to Penrose twistors. Starting with the
correspondence between the doubleton representations, homogeneous
functions on projective twistor space and on-shell generalized Weyl curvature $SL(2,\mathbb
C)$ spinors and their low-spin
counterparts, we study in the similar way the correspondence
between the massless representations, homogeneous functions on
ambitwistor space and, via the Penrose transform, with the gauge
fields on Minkowski boundary of $AdS_5$. The possibilities of
reconstructing massless fields on $AdS_5$ and some applications
are also discussed.
\end{abstract}

\section{Introduction}

The problem of characterization of irreducible
unitary representations of $SU(2,2)$ has rather
long history in mathematical physics (see \cite{Mack'77}
and \cite{Knapp'82}, where references to earlier literature can be
found). In the mid 80-s the interest in positive energy unitary representations of
corresponding supergroups $SU(2,2|N)$ \cite{GM'85}, \cite{Dobrev} was stimulated mainly by the
development of supersymmetry and supergravity,
in particular the necessity to describe the spectrum of $D=10$ IIB supergravity compactified on
$AdS_5\times S^5$ that was shown \cite{GM'85}, \cite{KRvN} to be given by the infinite-tower of massless and massive representations of $SU(2,2|4)$. At the end of 90-s the interest in positive energy unitary representations of $SU(2,2|4)$ renewed (see, e.g.,
\cite{FFZaffaroni}, \cite{GMZ'98}) in the context of $AdS_5/CFT_4$
gauge/string duality, on both sides of which the $SU(2,2|4)$ supergroup constitutes finite-dimensional part of the infinite-dimensional symmetry related to integrable structure. More recently in the framework of vectorial $AdS_5/CFT_4$ duality it was shown \cite{Beccaria}
that the spectrum
of $D=5$ Vasiliev-type higher-spin gauge theories\footnote{Higher-spin theories involved include like already known ones \cite{Sezgin}, \cite{Vasiliev2001}, as well as those yet to be identified.}, dual to free $4d$ scalar,
spinor or Maxwell theories, is described by the infinite set of the positive energy unitary representations of $SU(2,2)$ corresponding to massless fields.

Lowest-weight (positive energy) irreducible unitary representations of $SU(2,2)$ and
$SU(2,2|N)$, as well as those of other (super)groups can be constructed
using quantized oscillators carrying fundamental representation
labels of the maximal compact subgroup \cite{GS'81}, \cite{BG'83}. In this approach (super)group generators are realized as bilinear combinations of oscillators and part of them, that contains only raising oscillators, is used to produce the whole representation by acting on associated lowest-weight vector annihilated by the lowering oscillators. In the case of $SU(2,2)$
and $SU(2,2|N)$ such $SU(2)\times SU(2)$ oscillators are
naturally combined into Penrose twistors and supertwistors
\cite{GMZ'98}, \cite{Kallosh}. (Super)twistor theory \cite{PR} in its turn
has long been known to provide interesting alternative to
traditional space-time description of $4d$ massless gauge fields
that after the construction of the
twistor-string models \cite{Witten03}, \cite{Berkovits} have got
significant attention and allowed to unveil remarkable features of
Yang-Mills/gravity amplitudes written in the spinorial form.

Superalgebras $SU(2,2|N)$ can be extended to infinite-dimensional
superalgebras \cite{Linetsky-AP'90}, \cite{Sezgin},
\cite{Vasiliev'01} that admit realizations in terms of above
mentioned quantized supertwistors (oscillators) that play the role
of auxiliary variables in the construction of $4d$ conformal
higher-spin theories \cite{FrLin-NPB91} and $5d$ higher-spin
theories \cite{Sezgin}, \cite{Vasiliev2001}, \cite{Vasiliev02}.
The spectrum of constituent gauge fields fits into the
representation of underlying higher-spin superalgebra and
decomposes into an infinite sum of massless representations of
$SU(2,2|N)$ with spins ranging from zero to infinity (see
\cite{Vasiliev04} and references therein). More recently it was
shown that these superalgebras admit the realization in terms of
deformed twistors as enveloping algebras of $SU(2,2|N)$
\cite{Govil}.

In view of the important role played by massless gauge fields in the bulk of
$AdS_5$ and on its $D=4$ Minkowski boundary, in this paper we examine the possibilities to
reconstruct bulk (Fang-)Fronsdal fields \cite{Fronsdal}, \cite{Fang-Fronsdal},\footnote{The issue of deriving non-linear extensions of the Fronsdal equations for bosonic fields on $AdS_4$ starting from Vasiliev equations was addressed recently in \cite{Skvortsov}.} starting from the corresponding
positive-energy irreducible unitary representations of $SU(2,2)$ and using the isomorphism between oscillators and twistors and the properties of
homogeneous functions on the (ambi)twistor space.

Positive energy irreducible unitary representations of $SU(2,2)$
can be labeled by positive (half-)integers $(E,j_1,j_2)$, where the $AdS_5$ energy
$E$ is the eigenvalue of the $u(1)$
generator and $j_{1,2}$ are the
representation labels of $SU(2)_{L(R)}$ factors of the maximal
compact subgroup $SU(2)_L\times SU(2)_R\times U(1)$.
%In general Lorentz-covariant fields on $AdS_5$ correspond to direct sums of
%positive-energy unitary representations of $SU(2,2)$ \cite{GMZ'98}.
Values of $AdS_5$ energy $E$ are bounded from below and the form of the bound depends on the spin $s=j_1+j_2$ and the representation
\cite{Mack'77}. The simplest doubleton representations
$(s+1,s,0)$ and $(s+1,0,s)$, the tensor products of which contain all other positive energy unitary irreducible representations and for which associated fields are localized on the $D=4$ Minkowski boundary
of $AdS_5$, saturate the bound
\beq\label{doubleton-bound}
E\geq s+1,\quad j_1j_2=0.
\eeq
For massless fields on $AdS_5$ the bound has the following form
\beq\label{massless-bound}
E\geq s+2,\quad j_1,j_2\geq0.
\eeq
Note that bounds
(\ref{doubleton-bound}), (\ref{massless-bound}) provide the simplest instances of generic
relation between the $AdS_{D}$ energy of the irreducible unitary
representation of $SO(2,D-1)$ and the labels of the corresponding
$SO(D-1)$ representation that was derived in \cite{Metsaev'94},
\cite{Metsaev'95} from the requirement of the positive
definiteness of scalar product in the Fock space of $SO(2,D-1)$
oscillators. Since these oscillators are the $SO(2,D-1)$ vectors,
part of them satisfy 'wrong sign' commutation relations so that
the norm of a state in such a Fock space can be positive, negative
or null. Then the condition of the norm positivity leads to the
above discussed energy bound in similarity with the derivation of
the values of critical dimension and intercept in the old
covariant quantization of (super)strings. On the contrary the
oscillator approach of Refs.~\cite{GS'81}, \cite{BG'83} applied to
the description of positive energy unitary representations of
$SU(2,2)$ relies on the introduction of (a number of copies of)
bosonic oscillators transforming in the fundamental representation
of the two $SU(2)_{L(R)}$ factors in the maximal compact subgroup of
$SU(2,2)$ that obey positive-definite commutation relations. This
approach thus resembles the light-cone gauge (super)string
quantization scheme.

In the next section taking doubletons as the simplest example we confront known (but generically considered independently) oscillator and twistor approaches to their description and then apply gained experience to $AdS_5$ massless fields starting from the corresponding representations. For them the oscillator description is also familiar starting from \cite{GM'85}. Twistor description naturally involves ambitwistors and we consider how the ambitwistor data applies to the construction of the (Fang-)Fronsdal fields.

\section{Doubletons and twistors}

Let us remind the relationship between the $SU(2)$ oscillators and Penrose twistors (see, e.g., \cite{GMZ'98}, \cite{Kallosh}). The former correspond to diagonalization of the 'metric'
\beq
H=\left(
\begin{array}{cc}
0 & I_{2\times2}\\[0.2cm]
I_{2\times2} & 0
\end{array}\right)
\eeq
used to contract indices of fundamental (twistor) and antifundamental (dual twistor) representations of $SU(2,2)$. The twistor is defined by its primary $\mu^\al$ and secondary $\bar u_{\dot\al}$ $SL(2,\mathbb C)$ spinor parts
\beq\label{twistor}
Z^{\boldsymbol{\al}}=\left(
\begin{array}{c}
\mu^\al \\
\bar u_{\dot\al}
\end{array}\right)
\eeq 
and similarly the dual twistor $\bar Z_{\boldsymbol{\al}}=(u_\al,\bar\mu^{\dot\al})$. Since in transition to the oscillator basis only $SU(2)$ covariance is retained, following \cite{Wess} we replace dotted indices of the $SL(2,\mathbb C)$ spinor parts of the twistor and its dual by undotted ones in the opposite position that  are identified with the $SU(2)$ spinor indices in accordance with the uniqueness of the $SU(2)$ spinor representation. Then the oscillator variables are defined by the linear combinations of the twistor components 
\beq 
a^{\al}=\frac{1}{\sqrt{2}}(-\mu^\al+\bar u^\al),\quad a_\al=\frac{1}{\sqrt{2}}(-\bar\mu_\al+u_\al) 
\eeq 
and 
\beq 
b_{\al}=\frac{1}{\sqrt{2}}(\bar\mu_\al+u_\al),\quad b^\al=\frac{1}{\sqrt{2}}(\mu^\al+\bar u^\al)
\eeq 
that can be viewed as a kind of the Bogulyubov transform (for further discussion on that point see, e.g., \cite{Vasiliev'01}). Inverse relations express spinor parts of the twistor and its dual via the oscillators\footnote{The oscillators with upper/lower indices are complex conjugate of each other but bars like in the case of $SL(2,\mathbb C)$ spinors with dotted indices are not placed conventionally. This does not cause a confusion since changing the position of indices is not required in constructing $SU(2,2)$ positive energy unitary representations.  } 
\beq\label{Bogolyubov1}
\mu^\al=\frac{1}{\sqrt{2}}(-a^\al+b^\al),\quad\bar u^{\al}=\frac{1}{\sqrt{2}}(a^\al+b^\al)
\eeq
and 
\beq\label{Bogolyubov2}
u_\al=\frac{1}{\sqrt{2}}(a_\al+b_\al),\quad\bar\mu_{\al}=\frac{1}{\sqrt{2}}(-a_\al+b_\al).
\eeq 

In quantum theory introduced above oscillators can be shown to satisfy commutation relations
\beq\label{oscillators}
[\hat a_\al,\hat a^\bt]=\de_\al^\bt,\quad [\hat b^\al,\hat b_\bt]=\de^\al_\bt 
\eeq 
that allows interpret $\hat a_\al$ and $\hat b^\al$ as annihilation operators and their conjugates $\hat a^\al=(\hat a_\al)^\dagger$, $\hat b_\al=(\hat b^\al)^\dagger$ as creation operators acting on the unitary vacuum $|0\rangle$. Important invariant -- the twistor norm then transforms into the difference between the occupation numbers of $b$- and $a$-oscillators
\beq\label{twistor-norm}
\bar ZZ\rightarrow-\hat N_a+\hat N_b,\quad \hat N_a=
\hat a^\al\hat a_\al,\quad\hat N_b=\hat b_\al\hat b^\al
\eeq
and $su(2,2)$ algebra relations can be realized by the bilinears of quantized $a$- and $b$-oscillators. 

Positive energy unitary representations one can build using just one copy of $a$- and $b$-oscillators (\ref{oscillators}) are called doubletons \cite{GM'85}, \cite{GMZ'98-2}. They
correspond to $4d$ massless fields 'living' on the Minkowski boundary of $AdS_5$.
The lowest-weight vectors corresponding to definite doubleton representations are constructed out of the product of creation $\hat a^\al$ or $\hat b_\al$ oscillators\footnote{Throughout the paper we
adhere to the notation that a number in round brackets
following an index stands for the group of indices equal to that
number that are symmetrized with unit weight. Accordingly a number in square brackets following an index
denotes the group of indices antisymmetrized with unit weight.}
\beq\label{doubleton-lwvs}
|\,\mbox{lwv}\rangle=\hat a^{\al(2s_L)}|\,0\rangle\quad\mbox{or}\quad\hat b_{\al(2s_R)}|\,0\rangle
\eeq
acting on the oscillator vacuum annihilated by the $\hat a_\al$ and $\hat b^\al$ operators.
The whole representation in the basis corresponding to the maximal compact subalgebra $SU(2)_{L}\times SU(2)_{R}\times U(1)$ is constructed by applying to (\ref{doubleton-lwvs}) the raising operators $\hat L_+\vp{L}^{\al}\vp{L}_\bt=\hat a^\al\hat b_\bt$. They commute with $-\hat N_a+\hat N_b$, thus in any representation fixed is the integer $-2s_L$ or $+2s_R$. The $SU(2)_L\times SU(2)_R$ labels $j_{1,2}$ are given by half the eigenvalues of $\hat N_a$ and $\hat N_b$ on the lowest-weight vectors and the energy equals $E=j_1+j_2+1=s_{L(R)}+1$.

In the twistor picture doubleton representations are described by homogeneous functions $f(Z)$ on the twistor space $\mathbb{PT}^\bullet$ or homogeneous functions $\tilde f(\bar Z)$ on the dual twistor space $\mathbb{PT}_\bullet$. Such a description is based on the quantized twistors
\beq
[\hat Z^{\boldsymbol{\al}},\hat{\bar Z}_{\boldsymbol{\bt}}]=\de^{\boldsymbol{\al}}_{\boldsymbol{\bt}}
\eeq
realization as the multiplication and differentiation operators
\beq\label{qtwistor-realization}
\hat Z^{\boldsymbol{\al}}\rightarrow Z^{\boldsymbol{\al}},\quad\hat{\bar Z}_{\boldsymbol{\al}}\rightarrow-\frac{\pt}{\pt Z^{\boldsymbol{\al}}}
\eeq
or vice versa
\beq\label{qtwistor-dual-realization}
\hat Z^{\boldsymbol{\al}}\rightarrow\frac{\pt}{\pt\vp{\hat Z}\bar Z_{\boldsymbol{\al}}},\quad\hat{\bar Z}_{\boldsymbol{\al}}\rightarrow\bar Z_{\boldsymbol{\al}}.
\eeq
Operator realization (\ref{qtwistor-realization}) is adapted to the action on the twistor space functions and (\ref{qtwistor-dual-realization}) -- on the dual twistor space functions.

In the twistor approach doubleton representations built upon the lowest-weight vectors (\ref{doubleton-lwvs}) are described by the homogeneous functions on the twistor space with the homogeneity degrees $2s_L-2$ or $-2s_R-2$ as follows from (\ref{twistor-norm}). The twistor helicity operator
\beq
\hat s=\frac14(\hat{\bar Z}\hat Z+\hat Z\hat{\bar Z})
\eeq
in the realization (\ref{qtwistor-realization}) acquires the form
\beq
\hat s=-\frac12Z\frac{\pt}{\pt Z}-1
\eeq
so that the function $f_{(2s_L-2)}(Z)$ homogeneous of degree $2s_L-2>-2$ corresponds to the field with negative helicity $-s_L$ that describes left-polarized massless particles, whereas the function $f_{(-2s_R-2)}(Z)$ homogeneous of degree $-2s_R-2<-2$ corresponds to the field with positive helicity $+s_R$ and right-polarized particles \cite{PR}. In the case of positive helicity fields
reconstruction of the on-shell curvatures (linearized Weyl
curvature $SL(2,\mathbb C)$ spinors) on the $D=4$ Minkowski space-time proceeds
using the contour integral representation
\beq\label{contour-integral} \bar W_{\dot\al(2s_R)}(x)=\int\bar
u_{\dot\bt}d\bar u^{\dot\bt}\,\bar u_{\dot\al_1}\cdots\bar
u_{\dot\al_{2s_R}}\, f_{(-2s_R-2)}(i\bar
u_{\dot\lambda}x^{\dot\lambda\lambda},\bar
u_{\dot\lambda}):\quad\pt^{\dot\al\al}\bar W_{\dot\al(2s_R)}(x)=0,
\eeq 
where the incidence relations $\mu^\al=i\bar
u_{\dot\al}\tilde x^{\dot\al\al}$ that express primary spinor part
of the twistor via the coordinates $\tilde
x^{\dot\al\al}=x^a\tilde\s_a^{\dot\al\al}$ of the (complexified
conformally compactified) $D=4$ Minkowski space-time are assumed
to hold. For the negative helicity fields cohomological arguments
suggest a description in terms of the spinor form $\Gamma_{\al\dot\al(2s_L-1)}(x)$ of linearized
Christoffel-type connections \cite{deWit'80} modulo the gauge
transformations. Accordingly Penrose transform of the dual-twistor-space function for negative helicity yields generalized Weyl curvature $SL(2,\mathbb C)$ spinor of opposite chirality $W_{\al(2s_L)}(x)$, while for positive helicity it produces Christoffel-type connection $\Gamma_{\dot\al\al(2s_R-1)}(x)$ (see Table~\ref{doublePT}).
\begin{table}
\renewcommand{\arraystretch}{1.25}
{\footnotesize \begin{center}\begin{tabular}{|c||c||c|c||c|c|}
\hline
irrep & helicity & hom. degree on $\mathbb{PT}^\bullet$ & $D=4$ field & hom. degree on $\mathbb{PT}_\bullet$ & $D=4$ field \\ \hline\hline
$(1-s,-s,0)$ & $s<0$ & $-2-2s$ & $\Gamma_{\al\dot\al(-2s-1)}(x)$ & $-2+2s$ & $W_{\al(-2s)}(x)$ \\ \hline
$(3,2,0)$ & -2 & 2 & $\Gamma_{\al\dot\al(3)}(x)$ & -6 & $W_{\al(4)}(x)$ \\ \hline
$(5/2,3/2,0)$ &-3/2 & 1 & $\Lambda_{\al\dot\al(2)}(x)$ & -5 & $\Psi_{\al(3)}(x)$ \\ \hline
$(2,1,0)$ &-1 & 0 & $A_{\al\dot\al}(x)$ & -4 & $f_{\al(2)}(x)$ \\ \hline
$(3/2,1/2,0)$ & -1/2 & -1 & $\lambda_\al(x)$ & -3 & $\lambda_{\al}(x)$ \\ \hline
$(2,0,0)$ & 0 & -2 & $\varphi(x)$ & -2 & $\varphi(x)$ \\ \hline
$(3/2,0,1/2)$ & 1/2 & -3 & $\lambda_{\dot\al}(x)$ & -1 & $\lambda_{\dot\al}(x)$ \\ \hline
$(2,0,1)$ & 1 & -4 & $f_{\dot\al(2)}(x)$ & 0 & $A_{\dot\al\al}(x)$ \\ \hline
$(5/2,0,3/2)$ & 3/2 & -5 & $\Psi_{\dot\al(3)}(x)$ & 1 & $\Lambda_{\dot\al\al(2)}(x)$ \\ \hline
$(3,0,2)$ & 2 & -6 & $W_{\dot\al(4)}(x)$ & 2 & $\Gamma_{\dot\al\al(3)}(x)$ \\ \hline
$(s+1,0,s)$ & $s>0$ & $-2-2s$ & $W_{\dot\al(2s)}(x)$ & $-2+2s$ & $\Gamma_{\dot\al\al(2s-1)}(x)$ \\ \hline
\end{tabular}
\caption{Description of $D=4$ massless gauge fields by homogeneous functions on $\mathbb{PT}^\bullet$ and $\mathbb{PT}_\bullet$}\label{doublePT}
\end{center} }
\end{table}
General relation between the homogeneity degrees of functions on
the twistor space $h_{\mathbb{PT}^\bullet}$ and on the dual
twistor space $h_{\mathbb{PT}_\bullet}$ that correspond to the field of 
helicity $s$ is
\beq
h_{\mathbb{PT}_\bullet}=-h_{\mathbb{PT}^\bullet}-4.
\eeq
The correspondence between
respective cohomology groups of homogeneous functions is known as
the twistor transform.

\section{Massless field representations and ambitwistors}

Description of the positive energy unitary irreducible representations associated with massless fields on $AdS_5$ necessitates introduction of two copies of $a$- and $b$-oscillators \cite{GM'85}, \cite{GMZ'98-2}
\beq\label{two-oscillators}
[\hat a_\al(p),\hat a^\bt(r)]=\de_{pr}\de_\al^\bt,\quad [\hat b^\al(p),\hat b_\bt(r)]=\de_{pr}\de^\al_\bt,\quad p,r=1,2.
\eeq
In the twistor approach this corresponds to dealing with two Penrose twistors and their duals. Associate in accordance with (\ref{Bogolyubov1}), (\ref{Bogolyubov2}) to the first set of oscillators the twistor $\hat Z^{\boldsymbol{\al}}$ and its dual $\hat{\bar Z}_{\boldsymbol{\al}}$, and analogously to the second set of oscillators -- another pair of twistors $\hat W^{\boldsymbol{\al}}=(\hat\nu^\al, \hat{\bar v}_{\dot\al})$ and $\hat{\bar W}_{\boldsymbol{\al}}=(\hat v_\al, \hat{\bar\nu}{}^{\dot\al})$. Similarly to the doubleton case, in any representation fixed is the difference between the occupation numbers of $a$- and $b$-oscillators of the first and the second sets
\beq
(-\hat N_a(1)+\hat N_b(1))|\,\mbox{lwv}\rangle=-2s_1|\,\mbox{lwv}\rangle,\quad (-\hat N_a(2)+\hat N_b(2))|\,\mbox{lwv}\rangle=2s_2|\,\mbox{lwv}\rangle.
\eeq
For the representations that correspond to massless fields on $AdS_5$ $s_{1,2}$ are positive (half-)integers modulo relabeling the oscillators.
So one is led to consider homogeneous functions on the ambitwistor space $\mathbb A$: 
\beq
\begin{array}{c}
Z\frac{\pt}{\pt Z}F_{(2s_1-2|2s_2-2)}(Z,\bar W)=(2s_1-2)F_{(2s_1-2|2s_2-2)}(Z,\bar W),\\[0.2cm]
\bar W\frac{\pt}{\pt\vp{\hat W}\bar W}F_{(2s_1-2|2s_2-2)}(Z,\bar W)=(2s_2-2)F_{(2s_1-2|2s_2-2)}(Z,\bar W).
\end{array}
\eeq
We have chosen $Z^{\boldsymbol{\al}}$ and $\bar W_{\boldsymbol{\al}}$ to parametrize $\mathbb A$, while the operators corresponding to $\bar Z_{\boldsymbol{\al}}$ and $W^{\boldsymbol{\al}}$ have been traded for the derivatives of $Z^{\boldsymbol{\al}}$ and $\bar W_{\boldsymbol{\al}}$ (cf.~(\ref{qtwistor-realization}), (\ref{qtwistor-dual-realization})).
The condition $\bar W_{\boldsymbol{\al}}Z^{\boldsymbol{\al}}=0$ on the ambitwistor space coordinates is imposed via the $\de$-function
\beq
F_{(2s_1-2|2s_2-2)}(Z,\bar W)=\de(\bar WZ)f_{(2s_1-1|2s_2-1)}(Z,\bar W).
\eeq
In the oscillator approach it translates into the constraint
\beq\label{orthogonality}
(\hat b_\al(2)\hat b^\al(1)-\hat a_\al(2)\hat a^\al(1))|\,\mbox{lwv}\rangle=0.
\eeq

For homogeneous functions $f_{(s-1|s-1)}(Z,\bar W)\equiv f_{(s-1)}(Z,\bar W)$ on $\mathbb A$ with $s$ non-negative integer details of the Penrose transform can be found, e.g., in \cite{Eastwood-AMS} or in Ref.~\cite{Mason}. On-shell of the incidence relations
\beq\label{incidence-relations}
\mu^\al=i\bar u_{\dot\al}\tilde x^{\dot\al\al},\quad\bar\nu^{\dot\al}=-i\tilde x^{\dot\al\al}v_\al
\eeq
the function $f_{(s-1)}$ satisfies
\beq
v^\al\bar u^{\dot\al}\partial_{\al\dot\al}f_{(s-1)}(\mu, \bar u, \bar\nu, v)=0.
\eeq
$f_{(s-1)}$ is cohomologically trivial since $H^1(\mathbb{CP}^1\times\mathbb{CP}^1,\mc O(s-1))=0$ implying that it can be globally defined as a polynomial of the respective degree in $\bar u^{\dot\al}$ and $v^\al$, the homogeneous coordinates on $\mathbb{CP}^1\times\mathbb{CP}^1$,
\beq
\label{Penrose-transform}
v^\al\bar u^{\dot\al}\pt_{\al\dot\al}f_{(s-1)}\Rightarrow v^{\al(s)}\bar u^{\dot\al(s)}\wt b_{\al(s)\dot\al(s)}(x),
\eeq
where the symmetric multispinor field $\wt b_{\al(s)\dot\al(s)}$ is defined modulo the gauge symmetry
\beq\label{tw-gauge-sym-4d}
\de\wt b_{\al(s)\dot\al(s)}(x)=\partial_{\al(1)\dot\al(1)}\wt\xi_{\al(s-1)\dot\al(s-1)}(x)
\eeq
with the symmetric multispinor parameter $\wt\xi_{\al(s-1)\dot\al(s-1)}$. In 4d vector notation it corresponds to symmetric traceless rank-$s$ tensor field $\wt b_{a(s)}(x)$\footnote{Tilde is used to indicate tracelessness of a tensor w.r.t. Minkowski metric.} and the gauge parameter is given by the symmetric traceless rank-$(s-1)$ tensor field $\wt\xi_{a(s-1)}(x)$ so that
\beq\label{tw-gauge-sym-4d-vec}
\de\wt b_{a(s)}(x)=\pt_{a(1)}\wt\xi_{a(s-1)}(x)-\frac{1}{s}\eta_{a(2)}\pt^c\wt\xi_{ca(s-2)}(x).
\eeq

The discussion of the last paragraph corresponds to the simplest case of totally symmetric massless bosonic fields. In the oscillator description the solution to the constraints
\beq
(-\hat N_a(1)+\hat N_b(1))|\,\mbox{lwv}\rangle=-s|\,\mbox{lwv}\rangle,\quad(-\hat N_a(2)+\hat N_b(2))|\,\mbox{lwv}\rangle=s|\,\mbox{lwv}\rangle,\quad s\geq0
\eeq
and (\ref{orthogonality}) is given by the lowest-weight vector (cf.~\cite{GMZ'98-2})
\beq
|\,\mbox{lwv}\rangle=\hat a(1)^{\al(s)}\hat b(2)_{\bt(s)}|\,0\rangle.
\eeq
Respective representation labels are $j_1=j_2=s/2$ and $E=s+2$.
%field is denoted $\Xi_{(s/2,s/2)}(s+2)$ there.

Generalizing the above consideration to the functions $f_{(s-1|s)}$ and $f_{(s|s-1)}$ on $\mathbb A$ allows to obtain also fermionic fields $\wt\psi_{\al(s+1)\dot\al(s)}(x)$ and $\wt\chi_{\al(s)\dot\al(s+1)}(x)$
defined modulo the gauge transformations
\beq
\de\wt\psi_{\al(s+1)\dot\al(s)}(x)=\partial_{\al(1)\dot\al(1)}\wt\varep_{\al(s)\dot\al(s-1)}(x)
\eeq
and
\beq
\de\wt\chi_{\al(s)\dot\al(s+1)}(x)=\partial_{\al(1)\dot\al(1)}\wt\epsilon_{\al(s-1)\dot\al(s)}(x).
\eeq
In vector form these fields are given by the totally symmetric $\s$-traceless tensor-spinors $\wt\psi_{a(s)\al}(x)$ and $\wt\chi_{a(s)\dot\al}(x)$, for which the gauge variations read
\beq\label{ferm-var-f1}
\de\wt\psi_{a(s)}(x)=\pt_{a(1)}\wt\varep_{a(s-1)}(x)+\frac{1}{2(s+1)}\s_{a(1)}\tilde\s^b\pt_b\wt\varep_{a(s-1)}(x)-\frac{1}{s+1}\eta_{a(2)}\pt^b\wt\varep_{ba(s-2)}(x)
\eeq
and
\beq\label{ferm-var-f2} 
\de\wt\chi_{a(s)}(x)=\pt_{a(1)}\wt\epsilon_{a(s-1)}(x)+\frac{1}{2(s+1)}\tilde\s_{a(1)}\s^b\pt_b\wt\epsilon_{a(s-1)}(x)-\frac{1}{s+1}\eta_{a(2)}\pt^b\wt\epsilon_{ba(s-2)}(x).
\eeq
Associated lowest-weight vectors in the oscillator approach
\beq
\hat a^{\al(s)}(1)\hat b_{\bt(s+1)}(2)|\,0\rangle,\quad\hat a^{\al(s+1)}(1)\hat b_{\bt(s)}(2)|\,0\rangle
\eeq
correspond to representations with $j_1=s/2, j_2=(s+1)/2$ and $j_1=(s+1)/2, j_2=s/2$ that both have $E=s+5/2$.

Generic ambitwistor functions $f_{(2s_1-1|2s_2-1)}(Z,\bar W)$ with $s_1,s_2>0$, $|s_1-s_2|>1/2$ give rise to multispinor bosonic fields $\wt b_{\al(2s_2)\dot\al(2s_1)}(x)$ if $s_1+s_2$ is integer, or, if $s_1+s_2$ is half-integer, fermionic fields $\wt\psi_{\al(2s_2)\dot\al(2s_1)}(x)$ ($s_2>s_1$) or $\wt\chi_{\al(2s_2)\dot\al(2s_1)}(x)$ ($s_1>s_2$). They are defined modulo the gauge freedom
\beq
\de\wt b_{\al(2s_2)\dot\al(2s_1)}(x)=\partial_{\al(1)\dot\al(1)}\wt\xi_{\al(2s_2-1)\dot\al(2s_1-1)}(x)
\eeq
and analogous transformations for $\wt\psi_{\al(2s_2)\dot\al(2s_1)}(x)$ and $\wt\chi_{\al(2s_2)\dot\al(2s_1)}(x)$. Respective lowest-weight vectors are
\beq\label{mixed-sym-lwv}
a^{\al(2s_1)}(1)b_{\bt(2s_2)}(2)|\,0\rangle
\eeq
and representation labels are $(s+2,s_1,s_2)$ with $s=s_1+s_2$. Together with the conjugate representations $(s+2,s_2,s_1)$ they make up completely traceless tensor fields $\wt b_{a(s_1+s_2)b(|s_1-s_2|)}(x)$ and tensor-bispinor fields $\wt\Psi_{a(s_1+s_2-1/2)b(|s_1-s_2|-1/2)}(x)$. These are mixed-symmetry fields associated with two-row Young tableaux. Above considered representations are the only massless ones. Other $SU(2,2)$ representations that can be constructed using the two sets of $a-$ and $b-$oscillators \cite{GMZ'98-2}, in particular those arising in the limit $s_1=0$ or $s_2=0$ of (\ref{mixed-sym-lwv}), correspond to the so called massive self-dual fields \cite{Metsaev02}.

Symmetric traceless bosonic fields (\ref{tw-gauge-sym-4d-vec}) and $\s-$traceless fermionic fields (\ref{ferm-var-f1}) and (\ref{ferm-var-f2}) can naturally be identified with bosonic and fermionic $D=4$ shadow fields \cite{Metsaev08}, \cite{Metsaev13}, and treated as boundary values of $AdS_5$ totally symmetric massless gauge fields corresponding to the non-normalizable solutions of the Dirichlet problem for the Fronsdal equations. $AdS/CFT$ adapted description of massless mixed-symmetry is much more involved.\footnote{To date available is light-cone gauge formulation of $AdS_5$ massless mixed-symmetry fields \cite{Metsaev02}, $SO(1,D-1)$-covariant formulation for three-cell hook-type Young diagram field \cite{Alkalaev}, as well as the ambient-space one \cite{Grigoriev}.}  These results are summarized in table~\ref{m0irreps}, where $a',b'=0,...,3,5$ are $D=5$ tangent-space vector indices.
\begin{table}
{\footnotesize
\begin{center}
\renewcommand{\arraystretch}{1.5}
\setlength{\tabcolsep}{4pt}
\begin{tabular}{|c|c|c|c|}
\hline
irrep $(E,j_1,j_2)$ & $f$ on $\mathbb A$ & $SL(2,\mathbb C)$ multispinors & $AdS_5$ field \\ \hline\hline
$(s+2, s/2, s/2)$  & $f_{(s-1)}$ & $\wt b_{\al(s)\dot\al(s)}$ & $\wt B_{a'(s)}$  \\
\hline
\begin{tabular}{c}$(s+5/2, s/2, (s+1)/2)$ \\ $\oplus(s+5/2, (s+1)/2, s/2)$ \end{tabular} &
\begin{tabular}{c} $f_{(s-1|s)}$ \\ $\oplus f_{(s|s-1)}$ \end{tabular} &
\begin{tabular}{c} $\wt\psi_{\al(s+1)\dot\al(s)}$ \\ $\oplus\wt\chi_{\al(s)\dot\al(s+1)}$ \end{tabular} &
$\wt\Psi_{a'(s)}$ \\
\hline
& & $s_1+s_2$ integer: & \\
& & $\wt b_{\al(2s_2)\dot\al(2s_1)}$ & $\wt B_{a'(s_1+s_2)b'(|s_1-s_2|)}$ \\
$(s_1+s_2+2, s_1, s_2)$ & $f_{(2s_1-1|2s_2-1)}$ & $\oplus\,\wt b_{\al(2s_1)\dot\al(2s_2)}$ & \\ \cline{3-4}
$\oplus(s_1+s_2+2, s_2, s_1)$ & $\oplus f_{(2s_2-1|2s_1-1)}$ & $s_1+s_2$ half-integer: & \\
$s_{1,2}>0$, $|s_1-s_2|>1/2$ & & $\wt\psi_{\al(2\textrm{max}\{s_1,s_2\})\dot\al(2\textrm{min}\{s_1,s_2\})}$ & $\wt\Psi_{a'(s_1+s_2-1/2)b'(|s_1-s_2|-1/2)}$ \\
& & $\oplus\wt\chi_{\al(2\textrm{min}\{s_1,s_2\})\dot\al(2\textrm{max}\{s_1,s_2\})}$ & \\ \hline
\end{tabular}
\caption{$SU(2,2)$ massless representations, respective ambitwistor functions and space-time fields}\label{m0irreps}
\end{center} }
\end{table}

The
direct reconstruction of $AdS_5$ fields requires modification of
the Penrose incidence relations (\ref{incidence-relations}) to
accommodate the contribution of the fifth space-time coordinate.\footnote{For
the space-time of dimension $D=6$, for which the space-time
coordinate matrices span the space of $4\times4$ antisymmetric
matrices, natural generalization of Penrose twistors and on-shell  contour
integral relations does exist \cite{SWolf}, \cite{MRETCh}.} It is possible
to unify secondary spinor parts of ambitwistors $v^\al$ and $\bar
u_{\dot\al}$ into four-component $D=5$ spinor
\beq\label{5d-spinor} U_{\un\al}=\left(\begin{array}{c}
v_\al \\[0.2cm] \bar u^{\dot\al}
\end{array}
\right)
\eeq
and combine it with the Dirac conjugate into $SU(2)-$symplectic Majorana spinor. Such an $SU(2)-$symplectic Majorana spinor can be treated as the $4\times2$ rectangular block of the $D=5$ spinor Lorentz-harmonic matrix. Then by generalizing the construction of \cite{DelducGS}, an integral of homogeneous functions that belong to definite $SU(2)$ irreducible representations over the spinor harmonics produces the spinor form of $D=5$ Yang-Mills, Rarita-Schwinger field strengths, linearized Weyl tensor and its higher-spin counterparts that satisfy Dirac-type equations \cite{U}. Alternatively $D=5$ symmetric traceless gauge fields can be obtained via the Penrose transform for $D=5$ ambitwistors \cite{Mason}.

\section{Conclusions}

In this note we discussed the correspondence between the positive
energy (lowest-weight) unitary irreducible representations of
$SU(2,2)$ and the space-time fields on $AdS_5$ taking as an
example doubleton and massless representations that saturate the
bounds $E=s+1$ and $E=s+2$ respectively. Isomorphism between
bosonic oscillators, that can be used to construct positive energy
unitary irreducible representations of $SU(2,2)$, and Penrose
twistors allows to establish
one-to-one correspondence between the doubleton
representations and homogeneous functions of the single argument on projective
twistor space (or dual projective twistor space) that via the Penrose transform yield on-shell linearized
Weyl curvature $SL(2,\mathbb C)$ spinors and their low-spin counterparts in 4
dimensions. We sought for possible extension of this
twistor description to the case of massless representations. Since
to construct massless representations it is necessary to use twice
more oscillators compared to the doubletons, their natural twistor
counterparts are ambitwistors. We have shown that Penrose
transform for homogeneous functions on ambitwistor space yields
shadow fields on $D=4$ Minkowski space-time that admit interpretation
as boundary values of the non-normalizable solutions to the
Dirichlet problem for (Fang-)Fronsdal equations for the
corresponding $AdS_5$ massless gauge fields. This establishes 
one-to-one correspondence between  homogeneous ambitwistor
functions and $SU(2,2)$ massless representations. The fact that one arrives at the boundary
values of $AdS_5$ massless gauge fields rather than the bulk
fields themselves is encoded in the form of the incidence
relations that take into account only  the contributions of
$D=4$ Minkowski space coordinates. Direct obtention of the $AdS_5$
massless gauge fields requires extension of Penrose incidence
relations to account for the contribution of the fifth space-time
coordinate.

Natural generalization of the results reported in this note is to
introduce supersymmetry and also consider massive representations
pertinent to the adjoint version of the $AdS_5/CFT_4$
correspondence. Potentially interesting applications
can be also in twistor-string theory. Ambitwistor string models
have already been used to reproduce $D=4$ Yang-Mills and Einstein
gravity tree amplitudes in \cite{Geyer}. It is tempting to
speculate that their appropriate ramifications can produce tree
amplitudes of $D=5$ gauge theories.

While finalizing this paper we learned about Ref.~\cite{Adamo} that to some extent overlaps with our results and clarifies some points that we discuss here.

\end{document}